# MIXTURE SURVIVAL MODELS METHODOLOGY: AN APPLICATION TO CANCER IMMUNOTHERAPY ASSESSMENT IN CLINICAL TRIALS

**Lizet Sánchez[1*], Patricia Lorenzo-Luaces[1], Claudia Fonte[2], Agustin Lage[1]**

[1] Clinical Research Division, Center of Molecular Immunology, Calle 216 esq 15. Atabey, Havana 11600, Cuba

[2] Faculty of Mathematics and Computing. University of Havana, San Lazaro y L, Havana, Cuba

*Corresponding author:

email: lsanchez@cim.sld.cu

# Abstract

Progress in immunotherapy revolutionized the treatment landscape for advanced lung cancer, raising survival expectations beyond those that were historically anticipated with this disease. In the present study, we describe the methods for the adjustment of mixture parametric models of two populations for survival analysis in the presence of long survivors. A methodology is proposed in several five steps: first, it is proposed to use the multimodality test to decide the number of subpopulations to be considered in the model, second to adjust simple parametric survival models and mixture distribution models, to estimate the parameters and to select the best model fitted the data, finally, to test the hypotheses to compare the effectiveness of immunotherapies in the context of randomized clinical trials. The methodology is illustrated with data from a clinical trial that evaluates the effectiveness of the therapeutic vaccine CIMAvaxEGF vs the best supportive care for the treatment of advanced lung cancer. The mixture survival model allows estimating the presence of a subpopulation of long survivors that is 44% for vaccinated patients. The differences between the treated and control group were significant in both subpopulations (population of short-term survival: $p = 0.001$, the population of long-term survival: $p = 0.0002$). For cancer therapies, where a proportion of patients achieves long-term control of the disease, the heterogeneity of the population must be taken into account. Mixture parametric models may be more suitable to detect the effectiveness of immunotherapies compared to standard models.

# 1. Introduction

Advances in immunotherapy have revolutionized the treatment landscape for advanced lung cancer, raising survival expectations beyond those historically anticipated with this disease [1, 2]. More important than the improvements in the average survival time is the presence of a long and stable plateau with a heavy censoring at the tail of the curve [3]. This means that a proportion of patients are still alive even after long follow-up, it suggests the disease is controlled in a subgroup of long-term survivors. This new phenomenon is not currently captured by the most commonly used statistical procedures for the survival analysis, which generally assume that all patients are equally susceptible to the event of interest [3]. As an alternative, mixture parametric survival models have been proposed to take into account the heterogeneous structure in the data analysis [4,5].

The finite mixture models have stayed widely used in many disciplines. For example, Nemec & Nemec [6] apply the mixture models of astronomy distributions to study the number of distinct stellar populations in the Milky Way; Cameron and Heckman [7] use these types of models to assess the effect of family history on educational performance; Kollu [8] refers to several applications in engineering, Neale [9] in genetics and Land [10] in the social sciences. These models can be easily applied to the data set in which two or more subpopulations are mixed. Because of its flexibility, these models have been received intense attention in recent years, both from a practical and theoretical point of view. A complete description of the theory of mixture distributions and their applications can be found in McLachlan and Peel [11] or in Titterington et al. [12]. Finite mixture models have considered different distributions according to the data and problems that are modeled. For example, West [13] and Richardson and Green [14] studied the fit and applications of the mixture of normal distributions, McLachlan and Peel [11] the mixture of t- students, Wiper and Rugginieri Gamma distributions [15], and Tsionas Weibull [16]. However, in the analysis of data from clinical trials, specifically those that evaluate the effect on survival is rarely used [17,18].

The present study proposes a methodology for the evaluation of the effect of therapies in the presence of long-term survivors in the context of clinical trials. We have structured the methodology as follows: in session 2.1 we describe the multimodality test, a necessary condition for the application of the methods proposed below; Session 2 .2 describes the parametric models of simple survival and mixture distributions; in session 2 .3 the methods for estimating parameters of the proposed models are explained; session 2.4 explains how to make the selection of the best model and finally in session 2.5 the hypothesis tests for comparing the effectiveness of therapies in the context of randomized clinical trials are detailed . Additionally, the proposed methodology is illustrated for the analysis of data from a clinical trial that evaluates the effectiveness of the therapeutic vaccine CIMAvaxEGF for the treatment of advanced lung cancer.

## 2. Methodology

### 2.1 Multimodality test

Before adjusting a mixture parametric model, we must prove the existence of multimodality in the data, which justifies the application of the model. In this work, we assume the approach proposed by Silverman [19] and adapted by Hall and York [20] for the case in which we want to test whether the true distribution of a variable in a population is unimodal or bi-modal. Formally, given a sample of a random variable with density function f (in our case the survival time), denoting by j the number of modes in f, the hypothesis to test will be considered:

H0: j ≤ k, against H1: j > k,

Where k ∈ Z + is the number of modes to contrast.

The test proposed by Silverman is based on the concept of a critical width. The critical width will be the smallest width h for which it is verified that the estimate of the density function has at most k modes. From this critical width, denoted by $h_k$, Silverman proposes to use this parameter as a contrast statistic. Therefore the null hypothesis will be rejected when $h_k$ is very "large", since this would mean that it is necessary to overstate the core estimate for get a k – modal structure. The critical width, $h_k$, would be the last value of h before one of the modes "separates" giving rise to an estimate with (k + 1) modes.

The R silvermantest package (available at http://cran.r-project.org/) has implemented this method, which facilitates its implementation in practice.

## 2.2 Survival parametric models

*Simple model*

Survival analysis is a branch of statistics for analyzing the expected duration of time until one or more events happen. For the subjects in which the event could not be observed at the end of the follow-up period, the final state of the patient is called "censored", since the actual duration of survival time cannot be known [21]. We define the random variable T as the time that elapses since the patient is diagnosed with the disease until he dies. The survival function is defined as S (T) = P (T> t) where 0 <t <1, which can be obtained as:

$$S(t) = \int_t^\infty f(u)du$$

Where f (*u*) is the density function. Simple models assume a unimodal distribution of survival. The survival function, in this case, can be estimated from the data assuming a parametric model. The most used are Exponential, Weibull, Log-logistic and Lognormal (Table 1).

**Table 1.** Density and survival functions for continuous parametric distributions: Exponential, Weibull, Log-logistic and Lognormal.

| Model | Density function $f(t)$ | Survival function $S(t)$ |
|---|---|---|
| Exponential | $\lambda.e^{-\lambda t}$ | $e^{-\lambda t}$ |
| Weibull | $k\lambda t^{k-1}e^{-\lambda t^k}$ | $e^{-\lambda t^k}$ |
| Log-logistic | $\dfrac{kt^{k-1}\lambda}{(1+\lambda t^k)^2}$ | $\dfrac{1}{1+\lambda t^k}$ |
| Lognormal | $\dfrac{1}{t\sqrt{2\pi}\sigma}e^{-\frac{(\log t - \mu)^2}{2\sigma^2}}$ | $1 - \Phi\left(\dfrac{\log t - \mu}{\sigma}\right)$ |

*Mixture parametric survival model*

A model with a mixture of distributions assumes the existence of two or more populations within the sample, for each of which, the random variable T follows a different distribution. This means that there is heterogeneity in patient survival and that individuals in each subpopulation have different risks of dying. In this study, we will work with two-component mixture models [18].

The density function for this model is given by:

$f(x; \psi) = \pi_1 f_1(x; \theta_1) + \pi_2 f_2(x; \theta_2)$

Where $f_1$ and $f_2$ are the density functions of two of the distributions given in table 1. The parameter π (population weight parameter) is such that 0 <π ≤ 1, $π_1$ can be interpreted as the proportion of one component or population in the model, and $π_2$ = 1 - $π_1$ the proportion of the other. The vector ψ includes the vectors π; $θ_1$ and $θ_2$.

## 2.3 Methods for parameter estimations

To estimate the parameters of the distribution of a simple model, the maximum likelihood method implemented by the Optimum function of R was used. In the mixture models, the parameters were estimated by this same method, using the EM (Expectation-Maximization) algorithm, which was implemented in R and it will be described later.

### *Likelihood function*

Let $x_1, x_2, ..., x_n$ be a countable set of values for the discrete random variable X and Pψ (x) = Pψ (X = x). For the correct values of the parameter vector ψ the function P allows us to find the probability that X takes each of the values $x_1$, $x_2$,…. Let's look at this same function from another point of view: let's say we know that the random variable X takes the given values and follows a distribution Pψ (x) for an unknown value of ψ, so let Φ be the space of the possible values of ψ, we can interpret Pψ (x) as a function of ψ given x. Seen in this way, the function Pψ (x) is known as the likelihood function and is denoted by Lx (ψ).

In general, when you have a continuous random variable, for uncensored data, the likelihood function is given by Lx (ψ) = $\prod_{i=1}^{n} f(xi; ψ)$, it is assumed that the experiments in which the values of X were found were independent and f (xi; ψ) (the density function corresponding to the distribution with which the variable is assumed) replaces Pψ (x). Thus, if we considered the mixture distribution model defined in the previous section, we obtain:

$$L_x(\psi) = \prod_{i=1}^{n}(\pi_1 f_1(x_i; \theta_1) + \pi_2 f_2(x_i; \theta_2))$$

A problem arises when considering censored data since the value of the variable for all the elements of the sample is not known exactly. However, taking into account the type of censorship existing in our data, this can be solved using the survival function instead of the density function (Since the information we have is that the individual survived at least a number of months, we take the probability of surviving *y* or more months instead of surviving exactly *y* months). Thus, the likelihood function is as follows:

$$L_x(\psi) = \prod_{i=1}^{n_1}(\pi_1 f_1(x_i; \theta_1) + \pi_2 f_2(x_i; \theta_2)) \prod_{j=1}^{n_2}(\pi_1 S_1(x_j; \theta_1) + \pi_2 S_2(x_j; \theta_2))$$

Where $n_1$ and $n_2$ are the number of not censored and censored individuals respectively.

Given a space of possible values θ for the parameter ψ, the maximum likelihood method consists of finding the maximum likelihood estimate $\hat{\psi} = \arg\max_{\psi \in \Phi} L_x(\psi)$. Taking into account the monotony of the logarithm function, it is sometimes convenient to use $L_x(\psi) = \ln L_x(\psi)$ (log-likelihood function) instead of the maximum likelihood function, to calculate the maximum likelihood estimate.

## *EM algorithm*

We now have to answer the question: How to maximize the log-likelihood function? For this, there are several methods among which are the Moments Method, the EM Method, and the Fisher Information Method. More information on these can be found in [22]. In this study, we will use the EM algorithm, which is one of the most used in the literature.

To make the notation less cumbersome while describing the algorithm, it is assumed that the data is not censored, otherwise they would have, x1, x2,..., xn1 uncensored data and y1, y2,..., and n2 censored data, and the density or survival function would be taken as appropriate, the rest being analogous.

Let x1, x2,..., xn the values that the variable takes in a sample of individuals. The information can be considered incomplete since it is unknown to which of the two existing populations, in which the random variable distributes differently, within the sample each individual belongs. To complete the information, the indicator variable zi, i = 1, 2,.., n is taken, where zi = 1 if x1 is given by f1 and zi = 0 otherwise. The system (zi, xi) will contain the complete information of the data. Initially, the zi values can be found as a random sample of Bernoulli distribution (α), where α is the weight parameter.

The EM algorithm is divided into two stages, stage E and stage M. Each of these is described below in the k + 1-th iteration. In stage E, the value of zi is estimated from the conditional expectation E (zi | xi), which, using the Bayes total probability formula, can be found as:

$$\hat{z}_i = E(\hat{z}_i | x_i) = \frac{\pi_1 f_1(x_i, \theta_1^{(k)})}{\pi_1 f_1(x_i, \theta_1^{(k)}) + \pi_2 f_2(x_i, \theta_2^{(k)})}$$

Where θ1 (k) and θ2 (k) are the parametric vectors of the distributions of each component, estimated in the k-th iteration.

In step M, the parameter values are maximized, completing the information with the ẑi found above.

$$\pi^{(k+1)} = \frac{1}{n}\sum_{i=1}^{n} \hat{z}_i$$

$$\theta_1^{(k+1)} = \arg\max_{\theta_1 \in \Phi} \sum_{i=1}^{n} \hat{z}_i \ln(f1(xi, \theta1))$$

$$\theta_2^{(k+1)} = \arg\max_{\theta_2 \in \Phi} \sum_{i=1}^{n} \hat{z}_i \ln(f1(xi, \theta2))$$

For a single component, knowing the individuals belonging to that component, the problem of finding the values of the parameters that maximize the likelihood function is no more difficult than estimating the parameters of a simple model. This was done, using the Optimum function of R.

The iterations of stages E and M of the algorithm are repeated until | lx (ψ (k + 1)) - lx (ψ (k)) | is less than a small value, specified previously. The EM algorithm satisfies that lx (ψ (k + 1)) ≥ lx (ψ (k)), this property is the fundamental reason for its convergence [21].

**2.4 Method for selecting the best model**

If the actual distribution (G) of the variable with density function g is known, to measure how our model f (x | θ̂) approximates, the Kullback-Leibler Information (K-L) is used. K-L is given by:

$$E_{G(X)}\left[\ln \frac{g(X)}{f(x,\theta)}\right] = E_{G(X)}[\ln g(X)] - E_{G(X)} \ln f(x, \hat{\theta})$$

Where the best model will be the one that differs to a lesser extent from the real one and, therefore, the value of the K-L Information is lower. Moreover, in most cases, as in ours, the actual distribution of the variable is unknown. However, note that the expression EG (X) [ln g (X)] is a common constant for all models, so that the best model will be that, such that the value of EG (X) ln f (x, θ̂) be higher. This value can be approximated by the log-likelihood function plus a bias b(θ̂). Depending on how this approach is taken, several Information Criteria are defined for the selection of the best model. This paper uses the Akaike Information Criterion (AIC), defined as −2ln (Lx (θ̂)) + 2k.

**2.5 Hypothesis testing to evaluate the effect of the treatment**

The Log-rank test is the most used method to compare the survival of groups. It has as a null hypothesis that there are no differences between the populations for the occurrence of an event at any time of the follow-up. The analysis is based on the time of the events (i.e. deaths) of each group, which is compared with the expected number of events if there were no differences between the groups. In this work, we have used this test for the direct comparison of the Treated and Control groups for the case in which the existence of a single homogeneous population is assumed, that is, when a simple survival model is assumed (model 0).

For the case in which we assume the existence of two subpopulations (model 1), we have taken, as a cut-off point, the intersection of the estimated density functions for each subpopulation. All patients with survival lower of this cut-off point have been classified in the short-term population and all patients above the cut-off are considered to belong to the long-term population.

A new immunotherapy (IT) treatment being evaluated may not have an effect on the parameters for both populations, or it could modify the proportion of patients, or the median survival of some subpopulation, or simultaneously both effects [23]. To evaluate the effect of immunotherapy we allowed the model parameters to depend on the treatment. Table 2 summarizes the assumptions of the model and hypothesis tests considered on the parameters to assess the effect of IT.

**Table 2**: Hypothesis tested to evaluate the effect of the treatment in the mixture parametric survival model

| Variant[A] | Hypothesis | Explanation | Mean structure | Mixing proportions |
|---|---|---|---|---|
| 2 | H0: $\beta_{1k}=0$, k=1,2 | There is an effect of the therapy on the parameters for median overall survival, but not on the mixing proportion parameters. | $\lambda_k = \beta_{0k} + \beta_{1k}*IT_1$, k=1,2 | $\pi_1 = logit(z_0)$, $\pi_2 = 1 - \pi_1$ |
| 3 | H0: $\alpha=0$ H0: $z_1=0$ | There is an effect of the therapy on the mixing proportion parameters, but not on the parameters for median overall survival. | $\lambda_2 = \lambda_1 + \alpha$ | $\pi_1 = logit(z_0 + z_1*IT)$ $\pi_2 = 1 - \pi_1$ |
| 4 | H0: $\beta_{1k}=0$ H0: $z_1=0$ | There is an effect of the therapy on the mixing proportion parameters and on the parameters for median overall survival. | $\lambda_{1k} = \beta_{0k} + \beta_{1k}*IT_1$ k=1,2 | $\pi_1 = logit(z_0 + z_1*IT)$ $\pi_2 = 1 - \pi_1$ |

[A] To incorporate the effect of treatment in this model, we considered three variants

Definition of terms:

$H_0$, the null hypothesis

IT is the variable for the treatment groups (Control: IT=0 and CIMAvaxEGF: $IT_1$=1)

k=1,2 represent the short- and long-term survival populations

$\pi_1$ and $\pi_2$ = the proportion of short-term and long-term survivors, respectively; (where k=1 or 2), with the restriction that $0 < \pi_k \leq 1$ and $\pi_1 + \pi_2 = 1$, are the mixing proportions for the kth population.

$\lambda_k$ is the scale parameters for the Weibull distribution

\* Note that if any of the $\beta_{1k}$ is not significant, the average overall survival for subpopulation k would not depend on the treatment and would be similar for treaties and controls.

# 3. Application to the data

## 3.1 Data description

Data from a randomized, multicenter, controlled clinical trial in patients with advanced NSCLC was used. The phase III trial evaluated the efficacy of CIMAvaxEGF, an EGF-based cancer vaccine compared with best supportive care (Control). The recruitment period was between July 5, 2006, and January 3, 2012. Patients in advanced stages (IIIb / IV) who received 4 to 6 cycles of platinum-based first-line chemotherapy and who had confirmation were included. of stable disease or an objective response at least 4 weeks before randomization. A total of 405 patients were randomized (2: 1) to vaccinated or control groups. The vaccination group (n = 270) received the CIMAvaxEGF vaccine plus the best supportive care while the control group (n = 135) only received the best supportive therapy. Both groups were balanced according to the prognostic factors of the disease. More details can be found in [24]. The main evaluation criterion of the trial was overall survival (OS).

## 3.2 Existence of bimodality

The p-values obtained for the different k considered in the Silverman test are shown in Fig 1. It can be seen that the unimodality hypothesis is rejected for both treaties (p = 0.01) and controls (p <0.001). The first value of the test that is not significant is k = 2 in both cases, which indicates that survival in the two groups has a bimodal distribution.

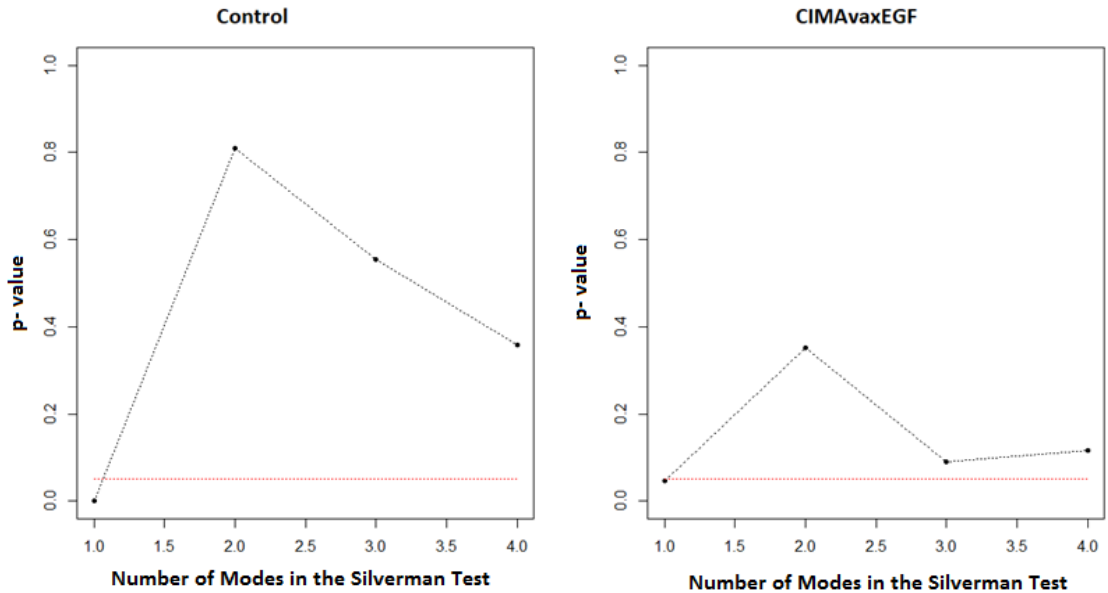

**Fig 1. P values according to the different k considered in the Silverman multimodality test for the survival distribution of the control groups and treated with CIMAvaxEGF. For k = 1 the values of p are less than 0.05, so the null hypothesis that the distribution is unimodal is rejected. The first non-significant value of p is obtained in both cases when k = 2 which is interpreted in this case as that the sample is bimodal.**

3.3 Models Fitting and best model selection

For the illustration of the proposed methodology, we have considered a Weibull distribution or a mixture of Weibull distributions. Table 3 shows the results of the adjustment of the different models and the estimated values of the median survival time and the proportions for the subpopulations. The two-component model resulted in a much better fit than the model assuming a homogeneous population (AIC: 2452 vs. 2783). The best fit is achieved with model 4 (AIC = 2448). This is the survival model that assumes the mixture of two distributions and that both parameters, median survival, and mixing proportion, depend on treatment. The components represent short-term and long-term survival subpopulations. A significant increase of 10% of patients with long-term survival is observed in the group treated with CIMAvaxEGF with respect to the control group (p = 0.005).

**Table 3**: Adjustment and estimates of model parameters assuming different hypotheses.

| Model | AIC | groups | Median1 | Median2 | π1 | π2 |
|---|---|---|---|---|---|---|
| 0 (1 component) | 2783 | - | 9.21 | | | |
| 1 (2 components) | 2452 | - | 7.28 | 27.29 | 0.55 | 0.45 |
| 2 (m~Treatment) | 2449 | Control | 6.28 | 26.24 | 0.62 | 0.37 |
| | | Treated | 8.35 | 34.95 | | |
| 3 (π~ Treatment) | 2449 | Control | 7.27 | 27.25 | 0.68 | 0.32 |
| | | Treated | | | 0.50 | 0.50 |
| 4 (m, π~ Treatment) | 2448 | Control | 6.69 | 26.01 | 0.66 | 0.34 |
| | | Treated | 7.91 | 30.78 | 0.56 | 0.44 |

### 3.4 Application to the evaluation of the effect of therapies

Fig 2 shows the survival curves for the short and long survival populations for both groups considering the best fit model (Model 4). The differences between the treated and control groups were significant in both subpopulations (short-lived population: p = 0.028; long-lived population: p = 0.0001).

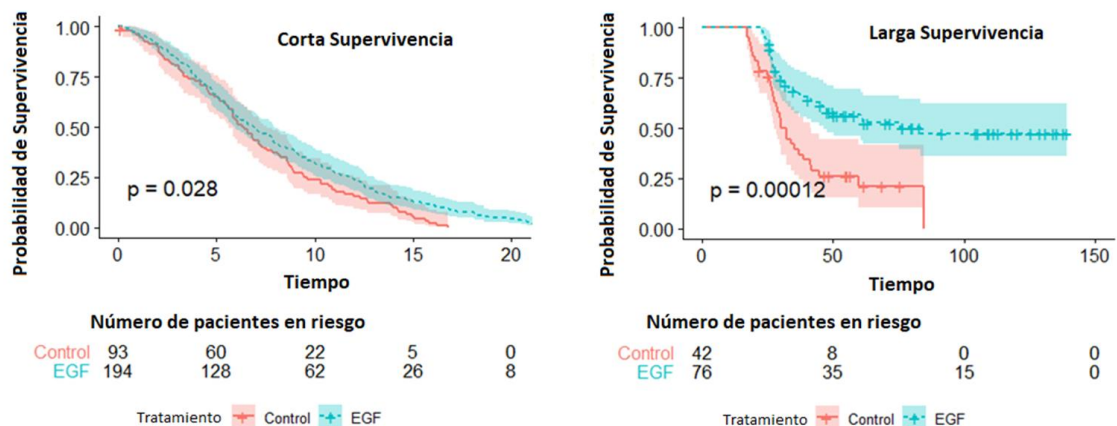

**Fig 2: Survival curves for short and long survival populations**

## 4. Discussion

In this work, a useful methodology was proposed to take into account the heterogeneity in the survival data. The methodology was illustrated in the evaluation of the effect of a new immunotherapy. Specifically, for the study of the survival of lung cancer patients in advanced stages, it is shown that the use of mixture distribution models are more appropriate than the use of simple models. The results of the study can be interpreted from three points of view: the existence of subgroups with different risks of dying within the population of patients with advanced lung cancer regardless of the therapy used; the effects of the therapeutic vaccine CIMAvaxEGF in the subgroups of patients identified as short and long survival; and the adaptation of current statistical methods to face the transition from advanced cancer to a chronic disease.

It has been shown that therapies for certain types of cancer induce a subset of long-term survivors, such as melanoma [25], breast cancer [26] and multiple myeloma [27]. Specifically for advanced lung cancer, evidence has been found of patients at different risk of death in both population studies [29-30] and in clinical trials [31-33]. The identification of this finding before modern therapies suggests that some patients experience prolonged survival regardless of treatment [29]. In this location, the fraction of long survival has been estimated in about half of the participants in the clinical trial, while in population studies it is estimated at around 10% [30]. This difference is attributed to the restrictive selection criteria in clinical trials, which means that there is an overrepresentation of patients more likely to be long-term survivors.

It should be noted that in many of the studies that refer to the presence of long survivors, traditional survival methods that assume homogeneity in the sample are used, a stratified analysis is done according to groups defined by biomarkers or the group of patients is divided arbitrarily into of short and long survival. The biological characteristics that define the heterogeneity of the data or the patient populations that will evolve to a short or long survival are not always known in advance. Biomarkers are often not incorporated into the design of the clinical trial but arise in secondary analyzes that are performed afterward. The definition of long survivors varies widely for each cancer location and among studies reported in the literature. To more accurately measure the clinical benefits of therapies, the American Society of Clinical Oncology and the European Society of Medical Oncology have proposed the ASCO-VF and ESMO-MCBS scales respectively. Both proposals were recently amended to incorporate bonuses and adjustments that capture the tail of the survival curve [34,35]. However, a study that critically analyzes the application of these proposals in 107 clinical trials concludes that neither of the two proposals was consistent as a measure of the absolute survival benefit [36].

The models of the mixture of distributions presented here are an alternative to consider and can be a useful tool for reinterpreting data from clinical trials already carried out based on taking into account heterogeneity. In our study, we do a retrospective re-analysis of the data from a clinical trial, and certainly, the percentage of long survivors may be overestimated. However, due to the randomization process, this must have occurred to the same extent in the treated and control groups. More important than the estimation itself of the fraction of long survivors, is the proposal to apply the mixture distribution model to test hypotheses about the effect of immunotherapies in clinical trials. This is in our opinion the greatest novelty of this work.

The methodology proposed here has the advantage that heterogeneity assumes without requiring groups to be defined beforehand. This is illustrated for the case in which there are two populations, but can easily be extended to the case where there are multiple subpopulations. It is important to keep in mind that the estimation of the mixture fraction can be very sensitive to the parametric distribution chosen to work [28]. For this reason, the selection of the parametric distribution to model the observed data should be done carefully. McCullagh and Barry [37] proposed a model selection process algorithm and recommended adjusting different distributions to the data and selecting the best distribution using one of the available information criteria.

Once the existence of heterogeneity in the main variable that is evaluated in retrospective exploratory studies has been detected, confirmation of this finding is imposed through the design of new clinical trials. This is one of the current challenges facing statisticians working in immuno-oncology. Even more so when the implementation of sequential Bayesian designs with intermediate analyzes has become a standard practice in phase III clinical trials to allow early termination of the studies. The software available for sample size calculation, such as PROC POWER in SAS, the survival analysis module in PASS, the powerSurvEpi package in R, is based on standard proportional hazard (PH) models that are not appropriate for the design in the presence of a fraction of patients with long survival. Cai [38] has proposed a methodology, which he has applied and implemented in the NPHMC R package. In his work, he establishes the calculation of sample size in the case in which a group of patients reaches a cure with the treatment. However, in the case of clinical trials in patients in advanced stages not operated, patients are not cured, but the new treatments are focused on achieving stabilization of the disease. New studies could address the proposal of clinical trial designs that take into account the mix of different distributions or risks in the subpopulations of the study.

The implementation in the exposed case study confirmed that vaccination with CIMAvaxEGF prolongs the survival of patients with advanced lung cancer. Although in both populations, short and long survival, patients are benefited with immunotherapy, the benefit is markedly greater in the long-lived population. Subsequent studies could propose the characterization and identification of biomarkers predictive of successful treatment with CIMAvaxEGF of both subpopulations or the selection of patients with a high benefit with this immunotherapy.

Future research could focus on the reproducibility of the findings of the effect of other treatments for lung cancer on the mixture fraction or on the survival of subpopulations. Alternatively, the prospective identification of patients with long-term survival potential may allow the validation of biomarkers predictive of the effect of therapies and favor the use of more personalized therapeutic approaches. The phenomenon of multimodality and the effect of immunotherapies on other types of tumors could also be studied. Studies are needed to determine the host factors and molecular characteristics that influence the heterogeneity of the survival of cancer patients.

# Acknowledgement

We thank all the patients and staff of all the institutions that were involved in the clinical trial.

## Author Contributions

Conceived and designed the study: LS and AL. Implemented the methodology and analyzed the data: LS, PL and CF. Wrote the paper: LS, CF and AL. All authors participate in the interpretation of the data, critically revised subsequent drafts of the manuscript, and approved the final version.

## Financial Disclosure Statement

The authors received no specific funding for this work